# A new dose calculation system implemented in image domain - A multi-institutional study

Jiawei Fan [1,2,3,4*], Zhiqiang Liu [5*], Dong Yang [1,2,3,4], Jiazhou Wang [1,2,3,4], Kuo Men [5], Jianrong Dai [5], Weigang Hu [1,2,3,4]

1, Department of Radiation Oncology, Fudan University Shanghai Cancer Center, Shanghai 200032, China

2, Department of Oncology, Shanghai Medical College Fudan University, Shanghai 200032, China

3, Shanghai Clinical Research Center for Radiation Oncology

4, Shanghai Key Laboratory of Radiation Oncology, Shanghai 200032, China

5, National Cancer Center/National Clinical Research Center for Cancer/Cancer Hospital, Chinese Academy of Medical Sciences and Peking Union Medical College.

**Corresponding Author Name & Email Address**

Kuo Men, Email: menkuo126@126.com

Jianrong Dai, Email: dai_jianrong@cicams.ac.cn

Weigang Hu, Email: jackhuwg@hotmail.com

*Jiawei Fan and Zhiqiang Liu contributed equally.

## Abstract

**Background**

Modern intensity-modulated radiotherapy, aiming to deliver an accurate dose to the planning target volume while protecting the surrounding organs at risk, is regarded as the indispensable treatment for cancer in the clinic. An accurate dose calculation algorithm is crucial to speed up the optimization process in treatment planning. To improve the effectiveness of the dose calculation, deep learning methods are widely adopted to compute dose distributions from several variables mimicking the physical features required for dose calculation. However, these methods employed similar concepts and neural network inputs, and had only limited improvements to traditional dose calculation methods.

**Purpose**

In this work, we propose a new computing process, named DeepBEVdose, which is essentially distinct to the previous deep learning-based dose calculation methods.

**Methods**

We present a novel image-domain dose calculation algorithm to automatically compute dose distributions from the computer tomography images and radiation field fluence maps. Specifically, a novel beam's eye view calculation scheme is introduced to substitute the traditional trivial ray-tracing procedure that cannot be removed in the previously published dose calculation system. Under this new calculation scheme, a generic two-dimensional convolutional neural network and a minimum required inputs are qualified to perform the accurate dose calculation.

**Results**

We demonstrate the feasibility of this approach with datasets acquired from multi-institutions with two different tumor sites (nasopharynx and lung). The average pixel-wise difference, among all the testing cases (both internal and external), between the ground-truth and predicted result is within 3%. The average dose calculation time was only a few seconds for each beam with a regular computer equipped with one standard graphics processing unit.

**Conclusions**

We have developed a novel deep learning framework that effectively maps radiation fluences to dose distributions. The high accuracy and efficiency of the proposed approach indicate its potential for use in online adaptive plan optimization.

# 1. Introduction

Modern external beam radiation therapy, usually accomplished by using multiple beams originated from a linear accelerator, is a treatment for cancer in which the patient is exposed to high doses of ionizing radiation in order to control the growth of the tumor. A treatment plan with accurate and reliable dose calculation is a crucial evaluation factor affecting the radiation therapy prognosis [1]. Currently, the most used dose calculation methods are Monte Carlo (MC) method and kernel-based methods such as pencil beam convolution (PBC) algorithm, convolution/superposition (C/S) algorithms including analytical anisotropic algorithm (AAA) (Varian Medical System, Inc., Palo Alto, CA, USA) and collapse cone convolution (CCC) algorithm (Pinnacle, CMS XiO, etc.) [2–6]. MC method can result in very accurate dose calculation but at the cost of longer calculation time. Although the kernel-based methods, consisting of approximations and only partially handling the physical processes involved in the microscopic energy absorption, have less accuracy with respect to the MC simulation, they achieve enough precision for general clinical applications and are widely adopted in the modern treatment planning systems (TPS). However, potential larger deviations still occur in some severe inhomogeneities circumstances, particularly in those regions where charged particle equilibrium does not hold. For instance, some studies indicated that the kernel-based algorithms still produce deviations from the measurement by more than 5% under certain circumstances in lung cancer treatments [7,8]. Thus, a novel higher accuracy dose calculation algorithm with higher efficiency is required for modern precision radiation therapy.

Recently, deep learning (DL) methods are proved to be effective in various fields including organs at risk (OARs) segmentation [9], image translation and registration [10,11], automatic treatment planning [12–17]. They were also applied to address the insufficient speeds and accuracy of widely-used dose calculation algorithms to facilitate the development of novel dose calculation systems. A previous study proposed a dose calculation algorithm based on deep residual neural networks (18). This study first converted a two-dimensional (2D) fluence map into a three-dimensional (3D) volume by using ray tracing algorithm. Then an indirect relationship was built between a fluence map and its corresponding 3D dose distribution, utilizing a 3D U-Net like deep neural networks, by establishing a mapping between the converted 3D volume, computed tomography (CT) and 3D dose distribution. Xing et al. [18] demonstrated that a HDU-Net can be implemented to formulate the relationship between the fluence map projected ray tracing dose and the TPS calculated dose. Similarly, Kontaxis et al. [19] introduced five different ray tracing-based 3D volumes, including the mask of the segment, the distance from the source, the central beamline distance, the radiological depth and the volume density, to model physical features. These features are then input to a 3D U-Net and outputs the predicted 3D dose distribution. The overall performance of these DL-based dose calculation frameworks in terms of both accuracy and inference speed, makes it compelling for daily treatment planning applications.

It can be seen that ray tracing is an important and indispensable step to account for the tissue heterogeneity in both conventional kernel-based and new DL-based dose calculation methods. The radiation beam is traced through voxels of the patient's 3D CT to calculate the delivered dose in each voxel while taking factors like absorption, scatter, and other physical parameters into account. This process must be repeated for each source position (gantry angle) to determine the total dose in each voxel. This time-consuming step may not be suitable for the online adaptive workflows where faster dose calculations are needed. Moreover, as we progress to more advanced MRI-guided interventions, intrafraction replanning will require almost real-time and accurate dose calculations which is indeed one of the most challenging tasks in modern radiation therapy. The graphical processing unit (GPU) on modern graphics cards offers the possibility of accelerating arithmetically intensive tasks. The GPU versions of the ray tracing algorithm were proved to be capable of performing dose calculation in considerably less time [20–22]. However, high-performance computers are required to be equipped for these GPU-based algorithms, which will definitely increase the maintenance costs. In addition, the stability of these algorithms needs to be further inspected and improved, many difficulties still need to be overcome for widespread clinical applications.

Our research aims to propose an image domain deep learning framework based on common convolutional networks and prove its feasibility to directly predict the dose distribution from individual beam fluence map, without the assistance of the ray tracing procedure. The newly developed framework, named DeepBEVdose, is a deep learning pipeline that correlates the fluence map of each radiation beam to its corresponding dose distribution by encoding meaningful physical and geometric information into the learning process. For the proof-of-concept experiment, we trained and applied this framework on treatment plans from nasopharyngeal and lung patients previously treated in multiple clinics by Varian (Varian Medical System, Inc., Palo Alto, CA, USA) linear-accelerator. This deep learning assisted dose calculation framework aims to open a new perspective for dose calculation and demonstrate the feasibility to avoid the trivial ray tracing process in the conventional dose calculation methods.

## 2. Methods

The flowchart of the proposed DeepBEVdose strategy is shown in Fig. 1. The input to the model is the beam's eye view (BEV) projection of patient's CT volume, the distance parameter, individual beam fluence map and corresponding dose distribution. Instead of relying on a complex 3D network that indirectly maps the 3D dose volume from a 2D fluence map (as discussed in the previous study [23]), we have opted to utilize a 2D convolutional neural network (CNN) to establish a direct correlation between the 2D dose distribution and fluence map via image domain translation.

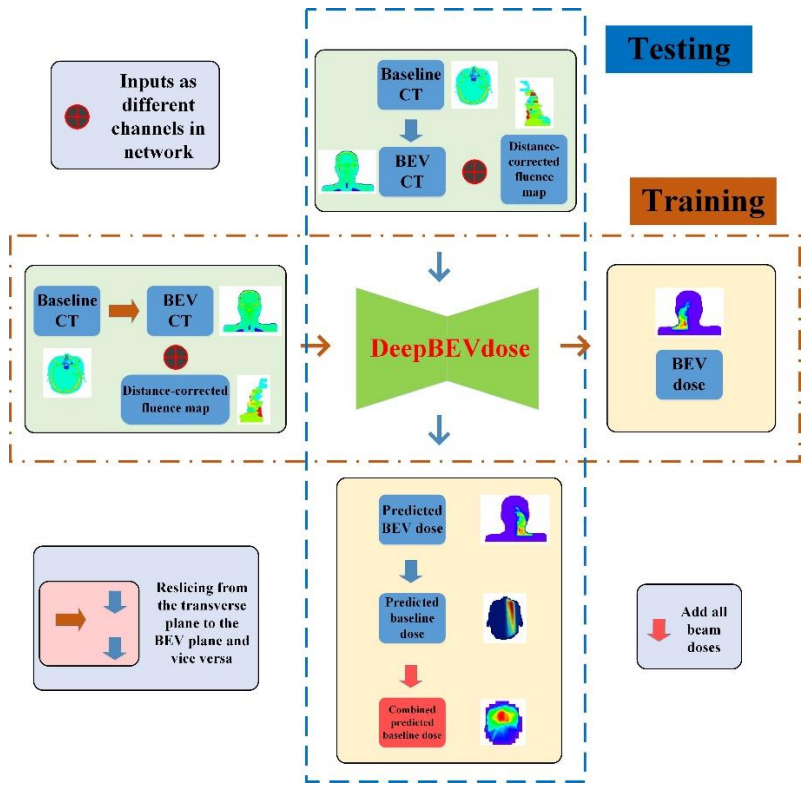

Fig.1. High-level overview of our proposed DeepBEVdose framework.

## 2.1. Dose calculation from BEV perspective

To accurately calculate the photon dose received by a patient undergoing clinical radiation therapy using X-rays from a linear accelerator (LINAC), an algorithm must be used to model the energy deposition pattern in the patient's tissues. This is done by taking into account the electron density information obtained from CT imaging of the patient. In this process, the LINAC head serves as the source of radiation while the patient's tissues and organs serve as the targets of interaction. Kernel-based dose calculation C/S algorithms used in commercial radiation therapy TPS utilize a kernel and ray tracing to model dose deposition ($r$) from an interaction at a primary interaction site ($r'$). The kernel represents the energy spread resulting from the interaction at the primary interaction site. The ray-tracing algorithms are used to project fluence map from the radiation source through the density representation of the patient. The absorbed dose can be computed as [24,25]:

$$D(r) = \iiint T(r') \times K(r - r')dV, \qquad (1)$$

where $T(r') = \frac{\mu}{\rho}\Psi(r')$, namely TERMA (total energy released per unit mass), is calculated from mass attenuation coefficient ($\frac{\mu}{\rho}$) and fluence map ($\Psi(r')$) involving trivial ray-tracing process, and $K(r - r')$ is density scaled point kernels, describing energy deposited at a radiological path length and angle away from primary interaction site, pre-calculating from complex MC simulations. The kernel is convolved with TERMA yielding the absorbed dose ($D(r)$).

Unlike the kernel-based algorithms, we proposed a new BEV calculation scheme absent of the trivial ray-tracing process. In the BEV perspective, each 2D dose slice has a different source-to-slice distance (SSD). Based on the inverse square law, the fluence map defined in the isocenter plane can be corrected to the fluence map at the specified slice. And the 2D slice dose are determined by this distance-corrected fluence map, and BEV CT. Thus, the proposed method trains a deep neural network by providing it with inputs such as the distance-corrected fluence map, the BEV CT, and the dose of a specific slice, which enables the network to establish a correlation between these variables. All physical processes, including photon attenuation and scatter, electron propagation, were assumed to be captured through model learning. In the testing stage, a baseline (in transverse plane) CT is first resliced from each BEV perspective to generate multiple 2D BEV slices. Each 2D slices corresponds to a distance-corrected fluence map. These are then fed as inputs to the trained neural network and outputs predicted 2D BEV dose for each slice. The predicted multiple 2D BEV doses are resampled to generate the final baseline 3D dose distribution.

## 2.2. DICOM Data analysis and extraction

In this study, we retrospectively collected a large cohort of patients containing nasopharyngeal and lung cancer cases from multiple clinics. The data were divided into three sets, with 180 patients allocated for training, 20 for validation, and 41 for testing (including 25 internal data and 16 external data). Notably, the training cases were sourced from single institute but the testing cases were from multiple clinics. Nasopharyngeal case presents complex target areas with multiple OARs that require careful consideration, making it one of the most challenging radiotherapy treatment sites. CT scans of lung cancer patients often reveal numerous cavities, and current DL-based methods frequently fall short in accurately predicting doses in these areas. Thus, careful studying these two datasets is particularly typical and demands a comprehensive approach for optimal prediction outcomes.

    All patients were planned in Pinnacle TPS with different prescription doses and beam orientations by intensity modulated static beam technique using 6 MV beam energy. The CCC algorithm was used as the ground-truth dose calculation engine,

with heterogeneity correction turned on. All treatment plans were designed and successfully delivered on the Varian Trilogy LINAC with a 120-leaf multi-leaf-collimator head.

For individual beam in each patient, the treatment plan was parsed to extract transverse CT volume, the fluence map and its corresponding 3D beam dose. The transverse CT volume and 3D beam doses underwent additional reslicing to align with the perspective of each BEV. We have determined that a single slice BEV dose, BEV CT, fluence map, and BEV distance from the slice to the radiation source combine to form a unique dataset. This dataset serves as the input for our network.

In the experiment, the image grid of fluence map was adjusted to 1 mm x 1 mm while the CT volume and 3D beam doses were rescaled to a uniform resolution of 1 mm x 1 mm x 5 mm. Before inputting the data into the deep neural network, they were transformed into numeric arrays with dimensions of 512 x 512. The centers of the images were maintained at their respective isocenters, and the grid was verified to safely encompass the patient's entire fluence map and dose distribution.

## 2.3. Network details

As mentioned, a single BEV slice of 3D beam dose is fully determined by the BEV CT, fluence map and the BEV distance from this slice to the radiation source. Thus, the problem of calculating the dose distribution can be approached as a mapping problem. Specifically, the aim is to use deep learning algorithms to map the BEV dose distribution to the fluence map, while taking into consideration of the various anatomical and geometric factors that may affect the radiation dose. Given the reliability and practicability of this mapping from the proposed BEV perspective, a common network structure that could be easily implemented in clinical application is sufficient. In our study, we utilized the residual blocks-based neural network operations that were originally proposed for predicting 3D dose distribution in nasopharyngeal cancer patients in the previous research [23]. We made further adjustments to the number and structure of the residual blocks in the network architecture, and carefully tuned the model parameters to address the specific challenges encountered in this study.

As evident from Fig. 2, individual CT slice and the BEV CT images together with the distance-corrected fluence map are employed as distinct inputs to the model. The model then generates the corresponding dose distribution as the output. The first half of the U-Net contains two hierarchies of five levels to reduce the feature sizes down to 16 x 16 and increase the filters to 512. This allows the model to learn both local and global features effectively. At each level, a residual convolutional block is used, followed by a variable number of residual identity blocks. The convolutional block applies a stride of size 2 x 2 to downsample the feature maps, while the identity blocks help to retain important features. On the right half of the network, two input feature maps are added and size of which is increased to 512 x 512 after a hierarchy of five levels. At each level, a residual transposed convolutional block is applied with a stride of size 2 x 2. This is followed by a varying number of residual identity blocks to further enhance the features. Skip connections are utilized in each level to maintain feature reusability from earlier layers, ensuring that features of the same dimensionality from previous layers are directly connected to the current layer, and helping to stabilize training and improve convergence. The final step in this right half of the network is a convolutional layer that generates a single channel as the final output. In each layer of the network, both convolutional and transposed convolutional operations are performed with kernel size of 3 x 3 or 1 x 1. Zero padding is applied to maintain the feature size during convolution or transposed convolution. Batch normalization is applied after both convolution and transposed convolution layers, followed by rectified linear unit (ReLU) activation function to introduce non-linearity and enhance feature representation. This helps in improving the performance and stability of the network during training and inference.

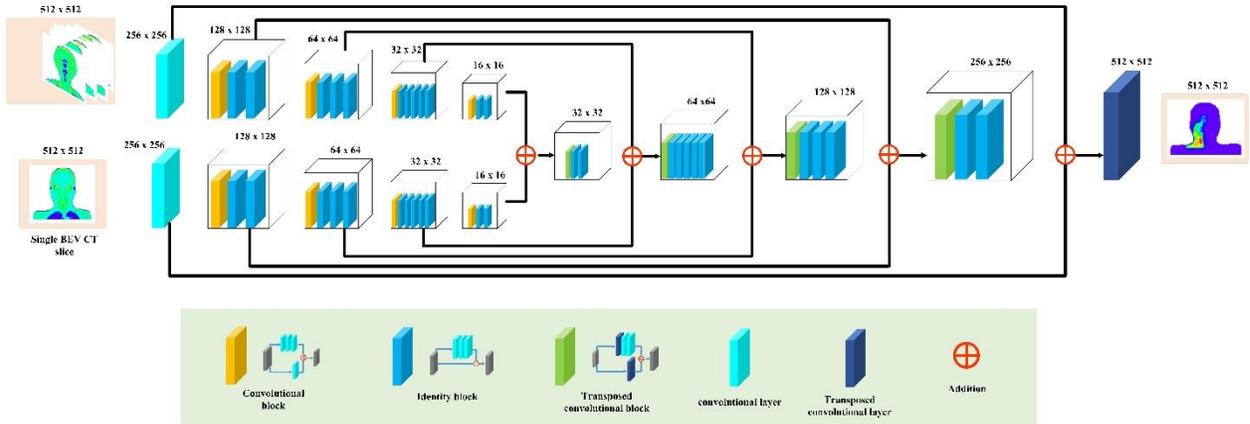

Fig.2. Illustration of the proposed UNet-like Network, including three residual blocks.

## 2.4. Training and testing

During training, extensive data augmentation techniques were effectively implemented to mitigate overfitting and improve the model's generalization performance. The initial learning rate was carefully tuned to 0.0001, and the widely used Adam optimizer was selected to optimize the model weights by minimizing the mean squared error (MSE) loss function. To prevent overfitting and ensure timely convergence, early stopping was employed based on model performance evaluation. The entire training process took approximately one week to obtain the final trained model. The deep neural network architecture was implemented using the PyTorch deep learning library, and the training was performed on a single NVIDIA 2080Ti GPU with 12 GB of dedicated memory, leveraging its powerful computational capabilities.

For the testing cases, we initially resampled each DL calculated BEV dose to the transverse plane. These resampled doses were then combined to generate the final combined dose for each individual beam. In order to evaluate the accuracy of the DL calculated dose distribution, we visually compared it with the dose distribution directly calculated from the TPS. Deviations from the prescription dose were computed between the DL and TPS calculated dose distributions to assess their differences. We also conducted a statistical analysis of the differences between the DL and TPS calculated dose distributions for all the testing cases, including calculating the pixel-wise dose differences and isodose volumes dice similarity coefficient (DSC). Additionally, we performed a dose-volume histogram (DVH) comparison of targets and OARs between the DL and TPS calculated results to further evaluate the model accuracy. Furthermore, we assessed and compared the quantitative dosimetric endpoints for the testing cases to provide a comprehensive evaluation of the DL calculated dose distribution. All these analyses provide us valuable insights into the performance of the proposed DL model.

## 3. Results

The trained model was applied to calculate the 41 testing cases, and the average calculation time for single case was within several seconds. Fig. 3 illustrates the TPS calculated doses (first row), the DL calculated doses (second row) and the dose difference (third row), from three different perspectives, loaded on one CT of the nasopharyngeal example patient. It is emphasized that the DL calculated doses are very close to the TPS calculated doses, the pixel-wise dose differences are within 2%. Fig. 4 (upper left) shows the DVH curves for the four internal testing patients in which the solid lines represent the TPS calculated dose and the dashed lines are the DL calculated doses. As depicted in these figures, the DL calculated DVH curves closely match those of the TPS calculated ones, indicating a clinically acceptable accuracy for the DL calculated results.

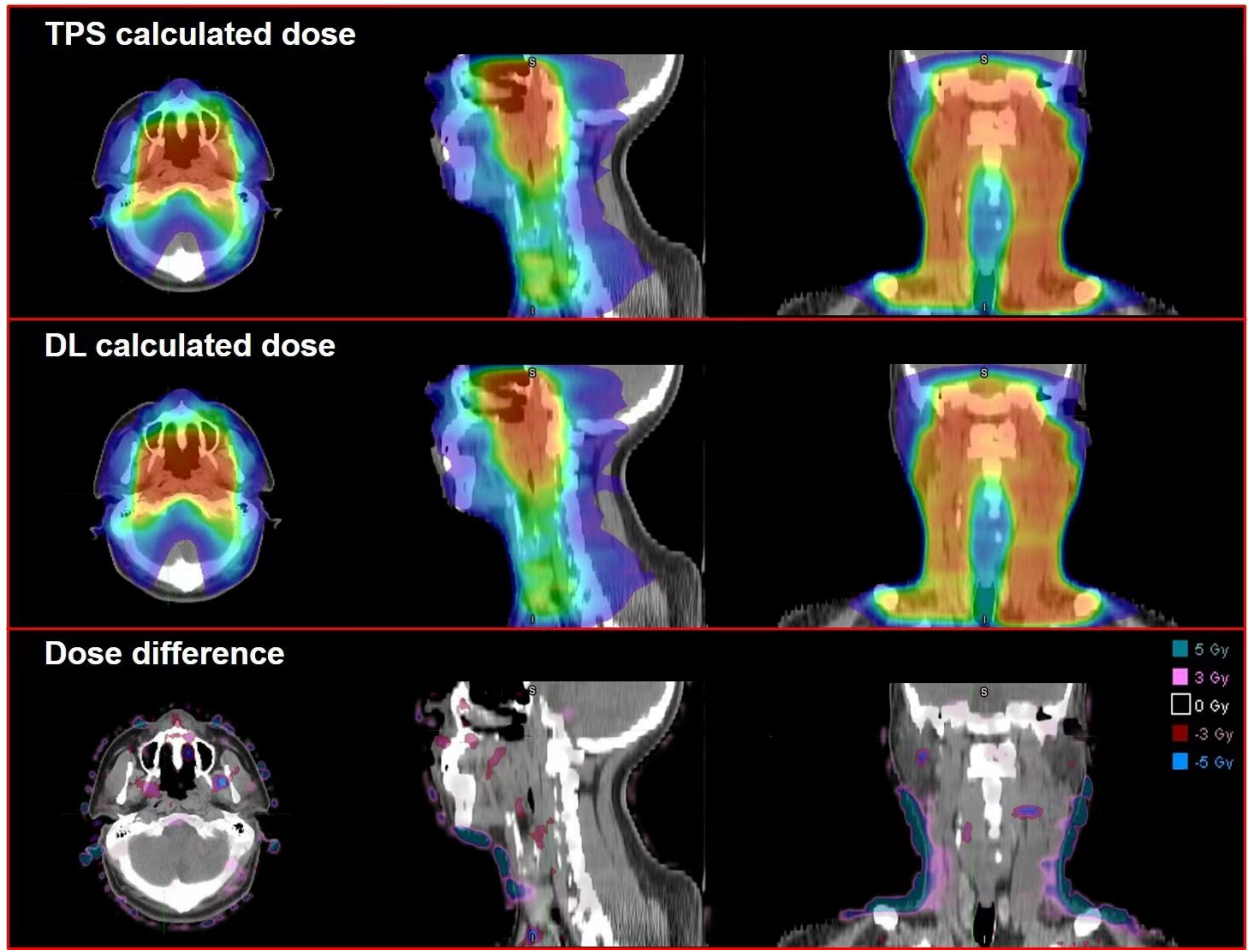

Fig.3. The TPS calculated doses (first row), the DL calculated doses (second row) and the dose differences (third row), from three different perspectives, loaded on one CT of the nasopharyngeal example patient.

Fig. 4 (bottom left) compares the isodose volumes Dice similarity coefficients between the TPS calculated and DL calculated dose distributions for all the internal testing patients. It can be seen that the proposed model predicts the dose band (DSC value from 0.85 to 1) fairly well. Fig. 5 (left) are box-whisker plots, demonstrating the median, 25% and 75% quartiles, minimum and maximum of the relative pixel-wise differences between DL and TPS calculated dose distributions within target and OAR regions, for all the internal testing cases. Note that most of the discrepancies are within 2% of the prescription doses which is in the clinically acceptable range.

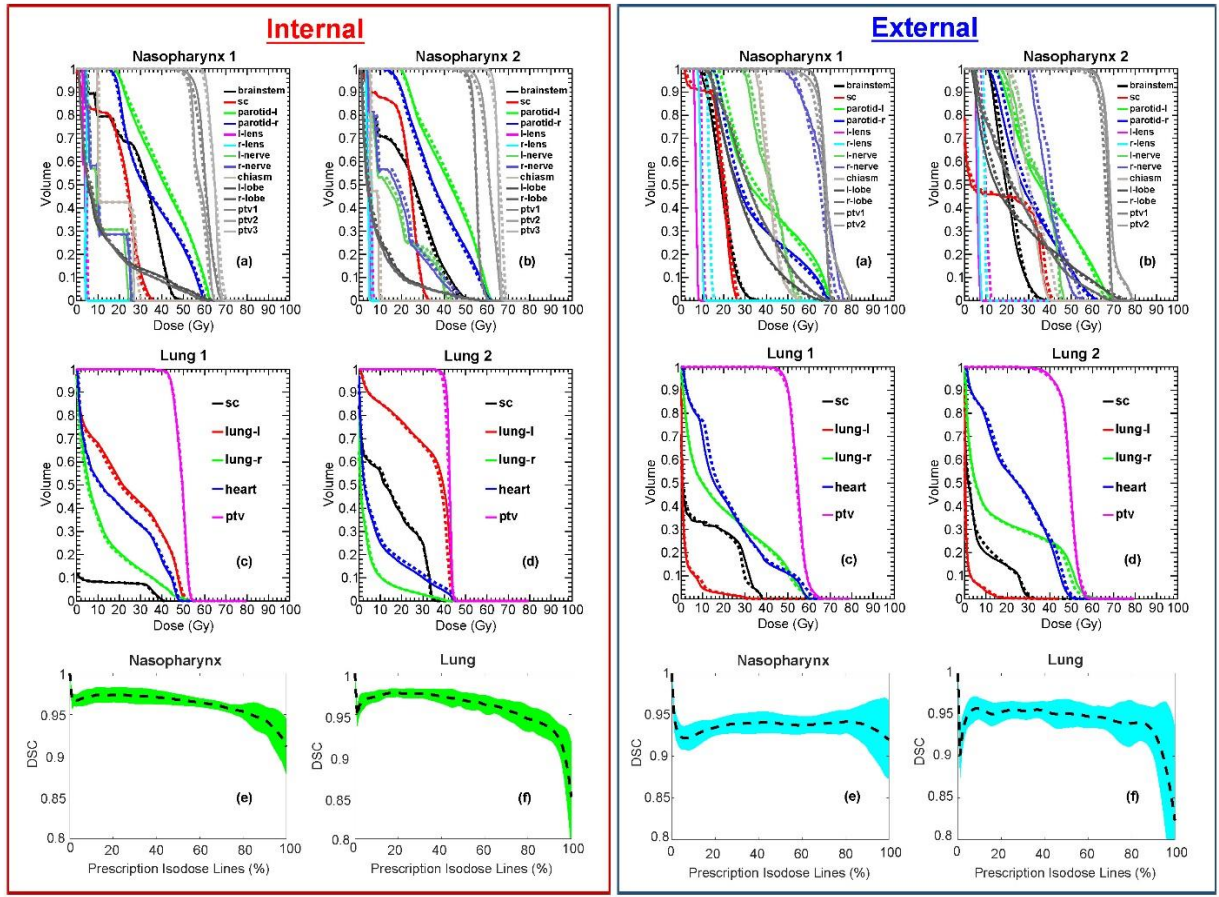

Fig.4. The DVH curves comparison for eight patients, the left is internal data and the right is external data. The solid lines represent the TPS calculated dose and the dashed lines are the DL calculated doses. The bottom four plots are DSC analysis results, comparing the isodose volumes between the TPS calculated and DL calculated dose distributions for all the internal and external testing patients. One standard deviation represented the error in the graphs.

Table 1 display the mean, standard deviation of the clinical evaluation criteria of both targets and OARs acquired from TPS and DL calculated dose distributions for all the internal and external testing patients. The Wilcoxon signed rank tests revealed that the DL calculated results did not statistically differ from the TPS calculated results for most of the clinical evaluation criteria. Therefore, the two dose distributions computed by TPS and DL are clinically identical.

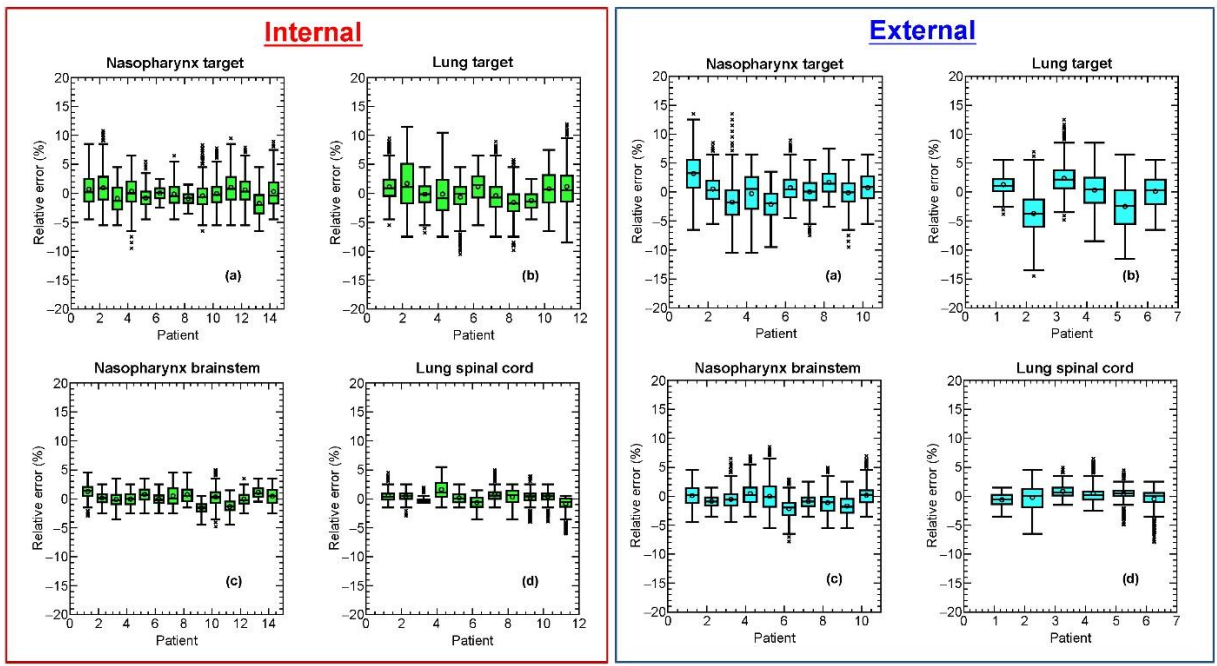

Fig.5. The box-whisker plots, demonstrating the median, 25% and 75% quartiles, minimum and maximum of the relative pixel-wise differences between TPS and DL calculated dose distributions within target and OAR regions. The left (a, b) is nasopharynx and lung target region in internal data, the left (c, d) is the brainstem of nasopharynx patient and spinal cord of lung patient in internal data. The right (a, b) is nasopharynx and lung target region in external data, the right (c, d) is the brainstem of nasopharynx patient and spinal cord of lung patient in external data.

Fig. 4 (right), Fig. 5 (right) and Table 1 display the similar results for patients obtained from the external clinic. The discrepancies of the external data are slight larger than the internal data, but still within the clinical acceptable range. It is evident that our model exhibits remarkable versatility and exceptional generalization performance.

Table 1. The mean, standard deviation of the clinical evaluation criteria of both targets and OARs acquired from TPS and DL calculated dose distributions for the internal and external testing data, respectively.

| Sites | Criterion | TPS | Prediction | Wilcoxon signed rank test | TPS | Prediction | Wilcoxon signed rank test |
|---|---|---|---|---|---|---|---|
| **Nasopharynx** | | | **Internal** | | | **External** | |
| Brainstem | Maximum | 50.17 ± 3.55 | 50.47 ± 2.83 | 0.78 | 44.60 ± 8.85 | 43.10 ± 7.91 | 0.71 |
| Spinal cord | Maximum | 38.83 ± 2.06 | 39.17 ± 2.22 | 0.65 | 32.60 ± 6.62 | 33.75 ± 5.93 | 0.70 |
| Right parotid | Mean | 33.31 ± 4.47 | 33.53 ± 3.86 | 0.88 | 30.27 ± 6.52 | 30.47 ± 6.61 | 0.95 |
| | $V_{30}$ (%) | 49.39 ± 12.64 | 49.72 ± 11.34 | 0.94 | 42.75 ± 9.20 | 43.26 ± 11.06 | 0.92 |
| Left parotid | Mean | 37.78 ± 3.91 | 38.31 ± 4.07 | 0.70 | 36.08 ± 4.82 | 36.52 ± 5.33 | 0.86 |
| | $V_{30}$ (%) | 59.21 ± 9.95 | 60.93 ± 10.40 | 0.62 | 54.66 ± 7.55 | 57.00 ± 9.45 | 0.57 |
| Right nerve | Maximum | 38.00 ± 17.48 | 39.67 ± 17.88 | 0.79 | 46.40 ± 19.66 | 46.30 ± 19.09 | 0.99 |
| Left nerve | Maximum | 39.08 ± 17.56 | 40.92 ± 17.93 | 0.77 | 45.55 ± 18.18 | 45.70 ± 18.48 | 0.99 |
| Right lobe | Maximum | 61.59 ± 5.38 | 63.22 ± 5.82 | 0.40 | 59.75 ± 16.48 | 59.95 ± 17.54 | 0.98 |
| Left lobe | Maximum | 62.69 ± 6.31 | 64.39 ± 6.44 | 0.44 | 60.90 ± 14.44 | 62.10 ± 16.01 | 0.87 |
| chiasm | Maximum | 30.03 ± 17.64 | 31.78 ± 18.35 | 0.78 | 44.55 ± 16.37 | 45.60 ± 16.12 | 0.89 |
| Right lens | Maximum | 5.42 ± 1.74 | 6.47 ± 2.17 | 0.13 | 6.75 ± 2.64 | 9.20 ± 3.75 | 0.13 |
| Left lens | Maximum | 5.58 ± 1.93 | 6.56 ± 2.22 | 0.18 | 6.25 ± 2.19 | 9.65 ± 3.37 | **0.02** |
| PTV1 | $D_{90}$ | 69.63 ± 3.32 | 70.10 ± 2.93 | 0.80 | 65.85 ± 11.72 | 66.77 ± 11.07 | 0.99 |
| PTV2 | $D_{90}$ | 59.730 ± 1.56 | 58.40 ± 0.61 | 0.32 | 60.58 ± 14.58 | 59.67 ± 13.91 | 0.89 |
| PTV3 | $D_{90}$ | 55.00 ± 3.15 | 54.07 ± 3.10 | 0.44 | | | |
| **Lung** | | | **Internal** | | | **External** | |
| Spinal cord | Maximum | 40.83 ± 3.05 | 41.00 ± 3.14 | 0.90 | 36.79 ± 2.02 | 37.29 ± 2.97 | 0.74 |
| Right lung | $V_{20}$ (%) | 26.06 ± 18.35 | 25.92 ± 18.31 | 0.99 | 17.98 ± 15.25 | 18.61 ± 15.25 | 0.94 |
| Left lung | $V_{20}$ (%) | 35.39 ± 26.72 | 35.04 ± 26.40 | 0.98 | 18.12 ± 18.65 | 17.37 ± 17.59 | 0.94 |
| Heart | Mean | 14.21 ± 9.78 | 14.69 ± 10.09 | 0.91 | 16.24 ± 9.02 | 16.72 ± 9.24 | 0.94 |
| PTV | $D_{90}$ | 49.65 ± 1.52 | 50.03 ± 0.84 | 0.75 | 53.99 ± 5.51 | 53.29 ± 5.10 | 0.82 |

## 4. Discussion

The accurate calculation of radiation dose plays a pivotal role in determining the efficacy of radiation therapy. However, a dilemma persists between efficiency and accuracy among the existing dose calculation engines in commonly used TPS. Fortunately, recent advancements in deep learning methods have opened up new possibilities for investigating novel dose calculation engines. The DL-based algorithms have shown promising results in achieving highly accurate dose calculations by leveraging the mapping between dose distribution images and radiation fluence images. These algorithms have demonstrated the capability to deliver dose calculations with both enhanced accuracy and efficiency, making them a compelling candidate for the next generation of dose calculation engines.

However, previous studies have not been able to eliminate the time-consuming ray tracing step, which translates geometric information into images, resulting in the need for substantial improvements before real clinical application. In this study, we focused on the feasibility of using deep learning techniques to develop a dose calculation engine without the ray tracing step. Our study demonstrated that by using the BEV perspective, we could establish the relationship between dose distribution and fluence map without the ray tracing step and successfully overcome the trade-off between computational speed and accuracy. The training of our deep learning model takes approximately one week, while the inference of the trained model takes only a few seconds. These times are no longer restricted by the time-consuming ray tracing process and can be further reduced by adopting more advanced GPUs. It's important to note that the proposed model can be trained with doses computed from any accurate dose calculation algorithms. In our future work, we expect even higher accuracy with real-time efficiency by using Monte Carlo calculated data as the model output.

We conducted extensive testing of our dose calculation engine, DeepBEVdose, on a diverse dataset consisting of 41 cases of nasopharyngeal and lung cancer patients from multiple institutions. Visual comparison of dose distributions and the pixel-wise differences, as shown in Fig. 3, confirmed the high accuracy of our proposed framework. To further validate our results, we performed statistical analysis of structural-specific DVH curves, isodose volumes Dice similarity, pixel-wise differences and clinical evaluation criteria, as presented in Fig. 4, 5 and Table 1. Our DL calculated results closely matched the ground-truth TPS calculated results, indicating the reliability of our model. Furthermore, we tested the generalizability of our model to external data, as shown in Fig. 4, 5 and Table 1. The predicted results, including dose distribution, DVH, and clinical key dose points, were still within clinically acceptable ranges, indicating the robustness of our model in different settings. Overall, most of the relative pixel-wise differences between TPS and DL calculated results were within 3% for all 41 testing cases, showing nearly perfect agreement. However, slightly larger discrepancies were observed in cavity and edge areas, which may be attributed to the inconsistency of dose calculation in these areas in the training data. This inconsistency arises from the inherent systematic errors of the dose calculation approximation model implemented in the TPS and could potentially be eliminated by using a Monte Carlo algorithm-based dose calculation engine. Meanwhile, the testing results of the external data are slightly worse, seen from Fig. 4, 5 and Table 1. Because different cancer centers may not have extremely identical dose restrictions or requirements for OAR and target areas, these differences will reflect in the external testing data and resulting in larger discrepancies.

Our current model is trained and tested on two specific cancer sites. However, we recognize the need to expand our model to include other clinical scenarios. In addition to the two cancer sites already covered, we plan to include additional cancer sites in our training data to ensure broader applicability of the model. Furthermore, we understand the importance of testing our model on different treatment machines, beam energies, and treatment techniques to ensure its versatility and robustness. To account for potential variations in different clinical scenarios, we will continuously evaluate and fine-tune our model. This may involve transfer learning, where the model learns from previously trained data and adapts to new scenarios, or re-training the model with updated data to ensure its accuracy and reliability. Our ultimate goal is to develop a comprehensive dose calculation system that can accurately predict radiation doses for a wide range of clinical scenarios.

The DeepBEVdose framework is based on a U-Net like architecture, leveraging the proven effectiveness of deep residual neural networks in various radiation therapy studies such as OAR segmentation, image translation, and automatic treatment planning. The simplicity and practicality of the residual neural network make it an ideal backbone for our framework. While more complex neural networks may also be competent, their integration into real clinical applications could increase complexity and may not meet expectations. Currently, our model is implemented in Python, a general-purpose interpreted programming language. Ongoing developments are focused on incorporating the DeepBEVdose framework into the TPS application programming interface (API), such as ESAPI, to enable seamless integration within TPS. We believe that the accuracy and convenience offered by our framework, as a secondary dose verification method, will assist physicians in making more informed decisions when evaluating treatment plans. As the proposed DL-based dose calculation engine differs from existing dose calculation algorithms, intensive developments are required to formulate a new generation of TPS that is entirely built upon deep learning techniques. This promising technology has the potential to increase the plan optimization speed and revolutionize traditional treatment planning approaches, and this will be our next step in this research direction.

The potential opportunities and challenges of utilizing the DeepBEVdose framework in MR-LINAC treatments are worth highlighting, particularly in the context of dose calculation for MRI-guided adaptive applications. These applications often require online dose calculations and treatment replanning in a short amount of time, while the patient is on the treatment couch. The proposed deep learning-based dose calculation engine has the potential to be highly useful in this regard. However, there is a need to improve the current framework to incorporate the effects of the transverse magnetic field of the MR-LINAC system into the dose calculation which would allow for a new dose calculation method based solely on MR images. This also presents an exciting opportunity to develop a new adaptive workflow that could enable real-time, adaptive treatment planning and delivery, lead to more precise and personalized radiation therapy.

## 5. Conclusions

We have developed a novel deep learning framework that effectively maps radiation fluences to dose distributions. Unlike traditional methods that rely on time-consuming ray tracing processes, our work marks the first successful instance of a dose calculation engine for radiation therapy that eliminates the need for ray tracing. Our results suggest that the proposed approach has the great potential to benefit various applications, including faster plan optimization, secondary dose calculation, and plan checking. Moving forward, we are committed to expanding the scope of our framework for broader applications and investigating its use in online adaptive workflows.

## Acknowledgements

This work is supported by the Special Research Fund for Central Universities, Peking Union Medical College, CAMS Innovation Fund for Medical Sciences (CIFMS) [2022-I2M-C&T-B-075], the Beijing Hope Run Special Fund of Cancer Foundation of China (LC2021B01), the National Natural Science Foundation of China (12375339), the National Key Research and Development Program of China (2022YFC2404603, 2022YFC2404600), the Shanghai Committee of Science and Technology Fund (21Y21900200), Xuhui District Artificial Intelligence Medical Hospital Cooperation Project (2021-012) and Key Clinical Specialty Project of Shanghai.

## Conflict of interest statement

The authors have no relevant conflicts of interest to disclose.